\documentstyle[aps,epsfig,multicol]{revtex}

\begin{document}
	
\bibliographystyle{unsrt}
\title{Bilateral symmetry breaking in a nonlinear Fabry-Perot 
cavity exhibiting optical tristability}
\author{Juan P. Torres, Jack Boyce and Raymond Y. Chiao}
\address{ Department of Physics, University of California, Berkeley, 
California 94720-7300}
\date{This version was produced on \today}

\maketitle

\begin{abstract}
We show the existence of a region in the parameter space that defines 
the field dynamics in a Fabry-Perot cylindrical cavity, where 
three output stable stationary states of the light are possible for a 
given localized incident field. Two of these states do not preserve the 
bilateral (i.e. left-right) symmetry of the entire system.  These 
broken-symmetry states are the high-transmission nonlinear modes of 
the system.  We also discuss how to excite these states.
\end{abstract}

\begin{multicols}{2}
Symmetries are at the backbone of physical theories.  In Hamiltonian 
systems, symmetries are related to conserved quantities through 
Noether's theorem, and they provide us with a powerful tool to 
understand Nature.  The left-right symmetry of the fundamental laws of 
Nature (apart from the nonconservation of parity for the weak 
interaction) is a well-known fact~\cite{lee}.  In particular, 
the electromagnetic interaction possesses this symmetry at the fundamental 
level. However, there is no reason why an invariance of the 
evolution equations should be an invariance of the stationary states 
of the system~\cite{coleman}. In fact, most of the time the state of 
really big systems does not have the symmetry of the laws which 
govern it~\cite{anderson}.  Here in a simple symmetrical dynamical 
system with a symmetric localized driving source, we have found an example of 
the dynamical breaking of a {\em discrete} fundamental 
symmetry of the system, namely, its left-right, or bilateral, 
symmetry.

Shortly after the proposal and demonstration of nondissipative optical 
bistability using Fabry-Perot cavities \cite{gibbs,felber}, the effect 
on the bistable behaviour of the transverse field amplitude profile of 
the fundamental mode of the cavity was taken into account 
\cite{marburger}. Nonlinearity and diffraction can excite several 
transverse modes of the cavity, and this will considerably modify the 
dynamics of the light~\cite{firth}.  For high finesse cavities, when only one 
longitudinal mode of the cavity is excited, the dynamics can be 
described with a single scalar wave 
equation~\cite{lugiato,haelterman1}, which facilitates numerical 
analysis and physical interpretation.  In this paper, we restrict 
ourselves to this regime.  Generally speaking, a great variety of 
light dynamics can be observed \cite{balkarei,jack}.  Concerning 
bistability, it has been shown that diffraction gives rise to 
transverse instability in one of the two branches of the plane wave 
bistable regime \cite{lugiato}.  This raises the question if 
bistability can exist at all when important transverse effects are taken into 
account. Some families of stationary solutions have been found 
\cite{haelterman2}, but their role in the dynamics of the light has 
not been discussed.  Here we consider new families of stationary 
solutions with broken symmetry which are relevant to the dynamics of 
the light that we numerically observed, and we discuss their stability 
and excitation.

We consider a cylindrical Fabry-Perot cavity~\cite{ivan} filled with a 
self-focusing Kerr nonlinear medium, $n=n_{0}€+n_{2}€|E|^{2}€$, where 
$n$ is the refractive index of the medium, $n_{0}€$ is the low-power 
refractive index, $n_{2€}>0$ is the nonlinear coefficient, and $E$ is 
the electric field inside the cavity (see Fig.  1).  The cavity is 
driven by a gaussian beam $E_{i}€$ with beam widths in the transverse 
$x$ and $y$ directions given by $w_{x}€$ and $w_{y}€$ ($1/e^{2}€$ half 
width intensity).  The concave mirror in the $y$-direction has a 
radius of curvature $R$, and the width $w_{y}€$ is chosen so that the 
incident beam matches the linear Hermite-Gaussian mode of the cavity 
in the $y$-direction.  Both mirrors have very high reflectivity 
$\cal{R}$, and the incident beam is assumed to have a Rayleigh range 
in the $x$ direction $z_{x}€=\pi w_{x}€^{2}€n_{0}€/ \lambda_{0}€$ much 
larger than the cavity length $L$.  $\lambda_{0}€$ is the light 
wavelength in vacuum.  Under these conditions, only a single 
longitudinal mode of the cavity is excited and the electric field $E$ 
has the structure $E(x,y,z,t) \propto \Psi(x,t) \exp(-y^{2}/w_{y}^{2}) 
\sin(k_{c}€ z) \exp(-i \omega €t) + c.c. \,$, where $\Psi$ is the 
normalized complex field envelope inside the cavity, $\omega$ is the 
frequency of the input beam $E_{i}€$, $t$ is the time and $k_{c}€$ is 
the resonance wavenumber of the linear cavity.  Only one transverse 
mode in the $y$ dimension is selected out by the use of the 
cylindrical cavity.  The field transmitted by the system is 
proportional to $\Psi$ \cite{haus}, which obeys the driven 
Ginzburg-Landau equation \cite{lugiato,haelterman1}

\begin{equation}
 \frac{\partial \Psi}{\partial \tau} = 
\frac{i}{2} \frac{\partial^{2}€ \Psi}{\partial \xi^{2}€}
- i \theta  \Psi 
+ i |\Psi|^2 \Psi
+ \Gamma( \Psi_{d}€- \Psi) 
\label{evolution}
\end{equation} 
where $\Psi_{d}€$ is the driving field, which is proportional to the 
incident field, $\xi=x/w_{x}€$ is the normalized $x$ coordinate 
and $\tau=t/t_{0}€$ is the normalized time.  The normalizing factor 
$t_{0}€$ is given by $t_{0}€= 2 n_{0}€z_{x}€ / c$, where $c$ is the 
velocity of light in vacuum.  The normalized cavity detuning $\theta$ 
is given by $\theta= 2 \Delta k z_{x}€$, where $\Delta k= k_{c}€-k€$ 
and the wavenumber $k$ is $k=n_{0}€\omega/c$.  The amplitude decay 
rate $\Gamma$ is $\Gamma=({\cal T} +\alpha L) z_{x}€ /L$, where ${\cal 
T} =1-{\cal R}$ is the transmissivity of each mirror, and $\alpha$ is 
the loss coefficient of the material filling the cavity.  The 
normalized amplitude $q$ of the driving field ($q\equiv\Psi_{d}(0)$), 
the normalized cavity detuning $\theta$, and the amplitude decay rate 
$\Gamma$ define the parameter space that determines the dynamics of 
the light in the Fabry-Perot etalon.

Let us consider a laser light that excites $^{85}€ {\rm Rb}$ atoms in 
their D$_{2}$ transition near $\lambda_{0}€=780$ nm to provide a 
resonantly-enhanced optical nonlinearity.  The mirror reflectivity is 
$\cal{R}$=0.995 and the cavity length $L=2$ mm.  The intracavity loss 
is $\alpha \simeq 0.1$ dB/cm and the linear refractive index is taken 
to be $n_{0}€ \simeq 1$.  A beam width $w_{x}€ \simeq 100 \,\mu$m 
corresponds to $\Gamma \simeq 0.15$, and $\theta \simeq 0.4$ 
corresponds to a cavity frequency detuning $\Delta f \simeq 240$ MHz.  
The Rayleigh range in the $x$ direction is $z_{x}€ \simeq 40$ mm.  The 
mode beam waist in the $y$ direction~\cite{milonni} is 
$w_{y}€=\sqrt{g^{2}€\lambda L / n_{0}€\pi (1-g)^{2}€}$, where 
$g=1-L/R$.  For $g=0.6$, we obtain $w_{y}€ \simeq 28 \,\mu$m, that can 
be achieved through an appropriate cylindrical mode-matching lens.  
The maximum nonlinear refractive index change $\Delta n$ is given by 
$\Delta n= \lambda |\Psi|^{2}€/4 \pi z_{x} A$, where $A=3 / 2 
\sqrt{8}$ is a mode overlap factor given by the field profile in the 
$y$ and $z$ directions.  For the parameters considered here, $\Psi 
\simeq 1$ corresponds to $\Delta n \simeq 3 \times 10^{-6}€$, which 
has been observed in a cylindrical Fabry-Perot experiment in rubidium 
vapor~\cite{jack}.

The time evolution of the field $\Psi$, as described by Eq.  
(\ref{evolution}), results from the interplay between different 
processes: transverse effects arising from the diffraction (second 
derivative) term in the wave equation, self-focusing due to the 
nonlinearity, energy input due to the driving field and energy loss 
due to the finite transmissivity of both mirrors.  The actual values 
of the parameters $\{ \Gamma, \theta, q \}$, will determine the 
influence on light dynamics of every one of the above mentioned 
processes.  For large beam width, we have $\Gamma$,$\theta \gg 1$, so 
transverse effects are expected to be negligible in this limit.  For 
small beam width, but large enough in order to be in the regime of 
validity of Eq.  (\ref{evolution}), the different competing processes 
depicted above are comparable.  This is the regime we are considering 
in this paper.  We will see that there is a region in parameter space 
where three possible output states with different broken and unbroken 
symmetries can be excited, and under appropriate circumstances, all 
stable output states can be excited.  In all our considerations we 
will restrict ourselves to the case $\theta > 0$, corresponding to a 
red-detuning of the laser from the cavity resonance.  Finally, we 
point out that the total energy inside the cavity and the transmitted 
beam power are both proportional to $I=\int_{-\infty}^{\infty}€ 
|\Psi|^{2}€ \, d\xi$, and this varies at a rate

\begin{equation}
\frac{d I}{d t}=-2 \Gamma \int_{-\infty}^{\infty}€ |\Psi|^{2}€ \,  d\xi 
 + \int_{-\infty}^{\infty}€ [ \Psi^{*}€ \Psi_{d}€ +\Psi \Psi_{d}^{*}€ ] 
 \, d\xi 
\label{energyflow}
\end{equation}  
Thus Eq. (\ref{evolution}) corresponds to a non-conservative system.

Let us first review the main results for the stability of plane wave 
solutions against transverse perturbations \cite{lugiato}.  In the 
plane wave limit, when the second derivative of the field in Eq.  
(\ref{evolution}) is neglected, the curve $|\Psi_{0}€|€-\Psi_{d}€$, 
where $\Psi_{0}€$ are the stationary plane wave solutions, is 
single-valued for $\theta < \sqrt{3} \Gamma$, whereas for $\theta > 
\sqrt{3} \Gamma$, is S-shaped and can lead to bistability.  Stationary 
plane-wave solutions are unstable against transverse perturbation when 
both conditions $|\Psi_{0}€|^{2}€ > \Gamma$ and $|\Psi_{0}€|^{2}€ > 
\theta/2$ are fulfilled.  Since the upper branch of the S-shaped curve 
$|\Psi_{0}€|-\Psi_{d}€$ always begins at $|\Psi_{0}€|^{2}€ > 2 
\,\theta /3$, it turns out that the upper branch is always unstable 
against transverse perturbation.  This raises the question whether or 
not bistability exists at all when one takes into account the 
transverse structure of the beams.

For that purpose, we first look for stationary solutions of Eq.  
(\ref{evolution}).  We set $\partial \Psi/\partial \tau =0$, and solve 
the corresponding ordinary differential equation with a Newton-Raphson 
scheme \cite{recipes}.  Note that contrary to travelling wave 
configurations, where stationary solutions are related to a nonlinear 
wavenumber shift \cite{akhmediev}, in the cavity configuration this is 
not the case, since the presence of the driving field makes Eq.  
(\ref{evolution}) no longer invariant under a global phase change.  We 
do not explore all the variety of stationary solutions that Eq.  
(\ref{evolution}) can support.  Instead, we restrict ourselves to the 
stationary solutions that will play a role in our discussion of 
bistability.  We plot in Fig.  2 families of stationary solutions 
corresponding to $\Gamma=0.15$ and $\theta=0.4$.  Stationary solutions 
plotted in Fig.  2 can be divided into two families.  The curve 
labeled $S$ corresponds to symmetric solutions $\Psi_{0}€(\xi)$, such 
that $\Psi_{0}€(\xi)=\Psi_{0}€(-\xi)$.  The curve labeled $B$ 
corresponds to to a pair of broken-symmetry solutions, $\Psi_{0}€(\xi) 
$ and $\Psi_{0}€(-\xi)$.  The curve $S$ has three branches defined by 
the sign of the slope of the curve in the figure, and we will refer to 
them as lower, middle, and upper branches.  Figure 3 shows the 
amplitude profiles of two stationary solutions, one with broken symmetry and 
the other with unbroken symmetry. Note the breaking of the left-right 
symmetry of the output field relative to that of the drive, for the 
broken-symmetry state.

To investigate the stability of the stationary solutions plotted in 
Fig. 2, we solve Eq.  (\ref{evolution}) with a split-step Fourier 
Transform algorithm \cite{agrawal}.  We take as input some selected 
slightly perturbed stationary solution 
$\Psi(\xi,\tau=0)=\Psi_{0}€(\xi) \left[ 1+v(\xi) \right]$, where 
$v(\xi)$ is a complex gaussian random variable for every value of the 
transverse coordinate with mean $\mu=0$ and typical deviation 
$\sigma=0.01$, for both the real and imaginary parts.  The 
perturbative noise will seed any instability present in the system.  
Since some of the stationary solutions have broken symmetry, in order 
to monitor the time evolution of the field, in addition to the peak 
amplitude, we will also follow the centroid $\bar{\xi}$, 
defined as
\begin{equation}
 \bar{\xi}(\tau)=\frac{\int_{-\infty}^{\infty}€ \xi |\Psi(\xi,\tau)|^{ 2}€}
          {\int_{-\infty}^{\infty}€ |\Psi(\xi,\tau)|^{2}€}.  
\label{centroid}
\end{equation}
For a symmetric beam, $\bar{\xi}(\tau)=0$.  After extensive numerical 
solutions, we find stable stationary solutions in the lower branch of 
curve $S$, and in curve $B$ below $q \simeq 1.48$.  This means that 
the outcome of the simulations are the corresponding stationary 
solutions $\Psi_{0}€$ in each case.  Stationary solutions 
corresponding to all other parts of curves $S$ and $B$ have been found 
to be unstable.  In some cases, the stability analysis leads to one or 
the other of the two corresponding stable stationary solutions for a 
given amplitude of the driving field, which one depending on the 
noise.  In other cases, the output is a time-varying pattern that 
depends on the particular value of the peak amplitude of the driving 
field.  In most cases, when the stationary solution of the lower 
branch of curve $S$ is not excited, the output of simulations always 
show that the beam does not preserve the bilateral symmetry of the 
driving field.  For values of the peak amplitude of the 
driving field $1.37 < q <1.48$, there are three stable stationary 
solutions.  The one corresponding to the lower branch of curve $S$, 
possesses an unbroken symmetry, and the transmissivity, defined as the 
ratio of the output beam power to the incident beam power, is low 
($\simeq 1 \%$).  The pair of solutions corresponding to curve $B$, 
one displaced to the left, the other to the right of the drive beam by 
equal amounts, possesses broken symmetry.  The transmissivity of this 
pair of solutions is remarkably high ($\simeq 75 \%$).  Thus the 
nonlinear Fabry-Perot cavity can be viewed as a nonlinear transmission 
device which selects out the broken-symmetry solutions as the 
high-transmission modes of the system.

Numerical simulations of Eq.  (\ref{evolution}) show that the 
symmetric state in the lower branch of curve $S$ can be excited after 
switching on the driving field with the appropriate amplitude of the 
driving field.  In order to excite the broken-symmetry  
states, a route to bistability has to be properly devised. Which one 
of the two broken-symmetry states is actually excited in the numerical 
simulations shown in Figs.  4 and 5, depends on round-off noise, 
and the method of excitation. One method is to tune the amplitude of the 
driving field in a step-like way to a value $q > 1.98$, so that a 
time-varying intermediate output state associated with curve $B$ is 
excited.  Then the amplitude of the driving field $q$ is reset to the 
value where the broken-symmetry state is stable. We plot an example 
of such route to tristability in Fig.  4.  Note the delay on the onset 
of the spontaneous symmetry breaking from the time the drive is 
switched down. The actual excitation of the 
broken-symmetry state depends on the characteristics of the 
intermediate unstable state that is excited, and on the duration of the 
round-trip time around the hysteresis loop.  Another way to excite the 
stable state in curve $B$, it is to switch on the driving field and 
then to slowly ramp it down.  Figure 5 show the actual excitation of 
both the symmetric and broken-symmetry states by this method.

We anticipate that these results, which we have obtained at the 
nonlinear classical field level, should be related to the scattering 
resonances of the underlying quantum field theory, in view of the 
correspondence principle.

This work has been supported by the Spanish Government under contract 
PB95-0768.  Juan P. Torres is grateful to the Spanish Government for 
funding of his sabbatical leave through the Secretar\'{\i}a de Estado 
de Universidades, Investigaci\'{o}n y Desarrollo.  Jack Boyce and 
Raymond Chiao acknowledge support of the ONR and the NSF.  We would also like to 
thank Eric Bolda, John Garrison, Morgan W. Mitchell, and Ewan Wright for very 
helpful discussions.

\newpage
\twocolumn

\begin{figure}
\centerline{\psfig{figure=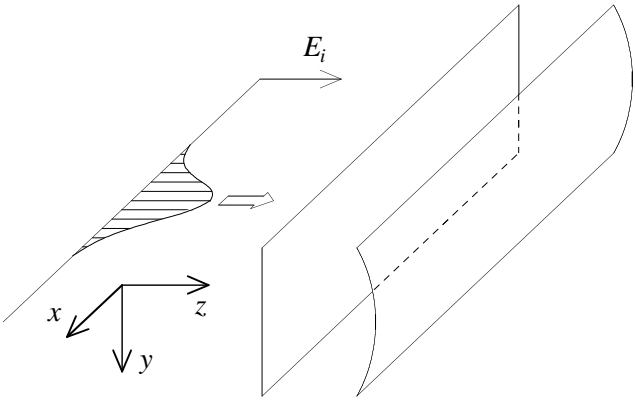,width=8.4cm}}
\caption{Geometry of the cylindrical Fabry-Perot cavity. }
\label{figure1}
\end{figure}

\begin{figure}
\centerline{\psfig{figure=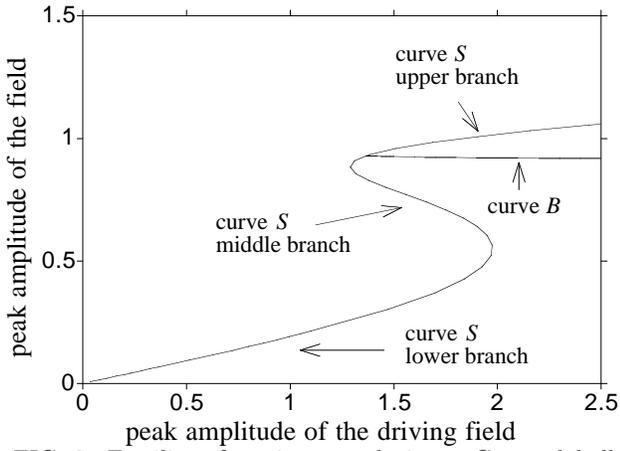,width=8.4cm}}
\caption{Families of stationary solutions. Curves labelled $S$ 
correspond to symmetric solutions, and curve labelled $B$ corresponds 
to broken-symmetry solutions. Parameters: $\Gamma=0.15$ and $\theta=0.4$.}
\label{figure2}
\end{figure}

\begin{figure}
\centerline{\psfig{figure=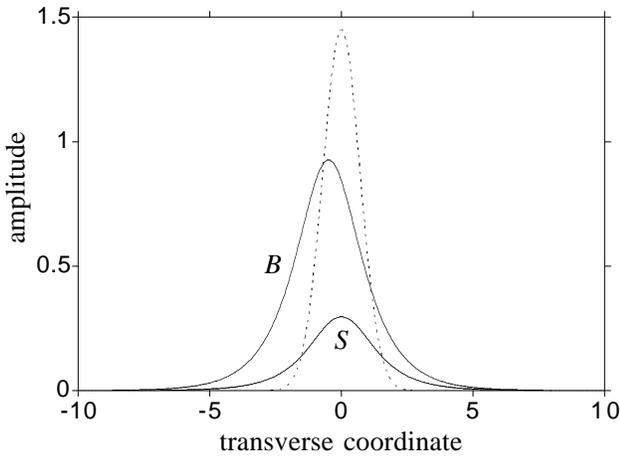,width=8.4cm}}
\caption{Amplitude profiel of stationary solutions from the $S$ and $B$ curves. 
Parameters: $\Gamma=0.15$, $\theta=0.4 $ and $q=1.45$. 
Solid line: amplitude profile of the field. Dotted line:
amplitude profile of the driving field. Note the breaking of left-right 
symmetry for $B$.}
\label{figure3}
\end{figure}

\newpage

\begin{figure}
\centerline{\psfig{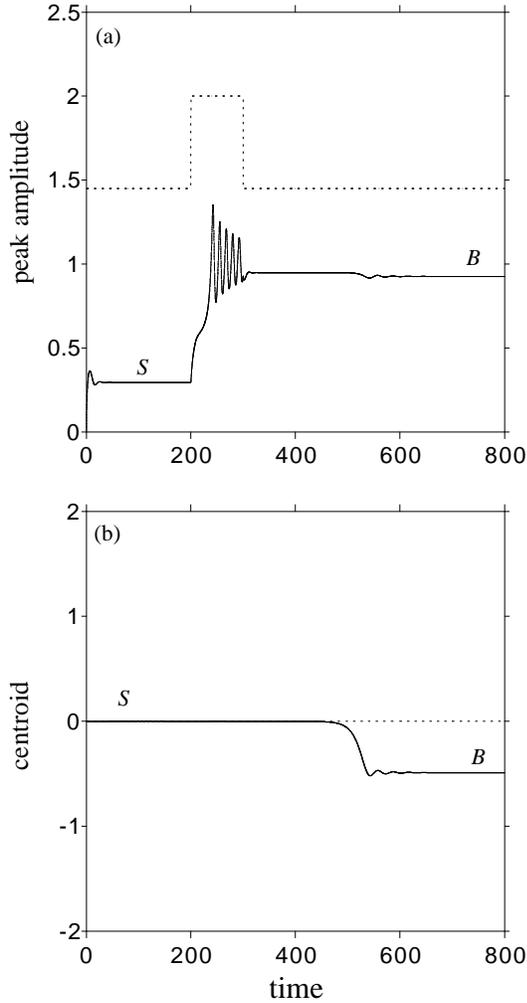}}
\caption{Step-like excitation of two stable states for $q=1.45$. 
(a) Peak amplitude and (b) centroid. 
Parameters: $\Gamma=0.15$, $\theta=0.4 $ and $q=1.45$.  $S$ 
corresponds to the excitation of the stationary solution in the lower 
branch of curve $S$, $B$ corresponds to the excitation of the 
stationary solution in curve $B$. Solid line: 
field. Dotted line: amplitude and centroid of the driving 
field.  Note the delay in the onset of  spontaneous symmetry 
breaking.}
\label{figure4}
\end{figure}

\begin{figure}
\centerline{\psfig{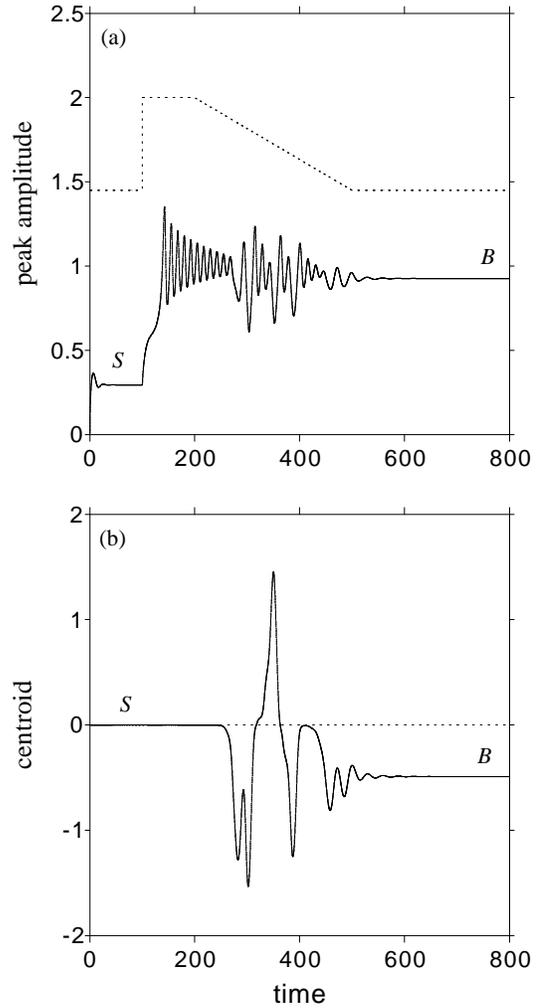}}
\caption{Ramp-like excitation of two stable states for $q=1.45$. 
(a) Peak amplitude and (b) centroid. 
Parameters: $\Gamma=0.15$, $\theta=0.4 $ and $q=1.45$. $S$ 
corresponds to the excitation of the stationary solution in the lower 
branch of curve $S$, and $B$ corresponds to the excitation of the 
stationary solution in curve $B$. Solid line: 
field. Dotted line: amplitude and  centroid of the driving field.}
\label{figure5}
\end{figure}

\end{multicols}

\end{document}